%% file: main.tex
\documentclass[twocolumn,nofootinbib,superscriptaddress,notitlepage,preprintnumbers]{revtex4-1}

\usepackage{orcidlink}

\usepackage{graphicx}
\usepackage{dcolumn}
\usepackage{bm}
\usepackage{hyperref}
\usepackage{bigints}
\usepackage{multirow}
\usepackage{enumitem}
\usepackage{upgreek}
\usepackage{url}
\usepackage{amsmath}
\usepackage{amssymb}
\usepackage{array}

\usepackage[english]{babel}
\usepackage{xcolor}
\hypersetup{
    colorlinks=true,
    linkcolor=[RGB]{252, 81, 133},
    citecolor=[RGB]{22, 121, 171},
    urlcolor=[RGB]{22, 121, 171}}

\usepackage{slashed,cancel}
\usepackage{xspace}

\usepackage[utf8]{inputenc}
	
\usepackage[T2A]{fontenc}
\usepackage[utf8]{inputenc}														
\usepackage{mathtext}

\usepackage{amsmath,amsfonts,amssymb,mathtools}	
\usepackage{icomma} 

\usepackage{babel,csquotes,xpatch}

\definecolor{jhg}{rgb}{0.984314,0.145098,0.462745}

\definecolor{jhg_edition}{rgb}{0.0862745, 0.47451, 0.670588}
\newcommand{\JE}[1]{\textcolor{jhg_edition}{#1}}

\definecolor{ac_edition}{rgb}{0.313725, 0.705882, 0.596078}

\input{custom_comands}

\begin{document}
\preprint{\hfill FTUV-24-0806.4693}

\title{\boldmath Heavy Neutral Lepton searches at an ICARUS-like detector using NuMI beam}

\newcommand{\PRL}
{Physical Research Laboratory, Ahmedabad, Gujarat, 380009, India}

\newcommand{\CERN}
{European Organization for Nuclear Research (CERN), 1211 Geneva 23, Switzerland}

\newcommand{\IFIC}
{Instituto de Física Corpuscular (IFIC), CSIC-UV, Carrer del Catedr\'atico Jos\'e Beltra\'an Martinez, 2, 46980 Paterna, Valencia, Spain}

\author{Animesh~Chatterjee
\orcidlink{0000-0002-2935-0958}}
\email[]{animesh.chatterjee@cern.ch} 
\affiliation{\PRL}
\affiliation{\CERN}

\author{Josu~Hernandez-Garcia
\orcidlink{0000-0003-0734-0879}}
\email[]{josu.hernandez@ific.uv.es}
\affiliation{\IFIC}

\author{Albert De Roeck
\orcidlink{0000-0002-9228-5271}}
\email[]{Albert.de.Roeck@cern.ch} 
\affiliation{\CERN}



\begin{abstract}
 The discovery of non-zero neutrino masses and mixings that the Standard Model (SM) cannot accommodate opens up the possibility of the existence of Heavy Neutral Leptons (HNLs). In minimal models, the HNL production and decay are controlled by SM interactions and the mixing between HNLs and the active neutrino and typically result in relatively long lifetimes if the masses are in the MeV-GeV range. We have studied the physics case and technical feasibility for a dedicated HNL search using the NuMI beam at an ICARUS-like detector. Our analysis demonstrates that the constraints on the mixing of the HNL as a function of its mass for an ICARUS-like detector with NuMI beam are highly competitive with the limits obtained from present experiments.
\end{abstract}

\maketitle
\flushbottom

\section{Introduction}
\label{sec:intro}

The Standard Model (SM) of particle physics, despite its very accurate predictions tested in a wide variety of experiments, is leaving several open questions unanswered. Among these are the absence of a dark matter candidate, the baryon asymmetry of the Universe (BAU), the hierarchy problem, the origin of neutrino masses, and flavor mixing observed in neutrino oscillation experiments. \\
Many extensions to the SM have been proposed to address these questions. One potential extension is the addition of right-handed neutrinos, which are singlets of the SM gauge group. This extension would allow for Yukawa couplings in the neutrino sector, in complete analogy to the other fermions of the SM. The singlet nature of the
right-handed neutrinos introduces a new energy
scale in the Lagrangian, the Majorana mass term, with
a dimensionful new parameter not related to the electroweak (EW) symmetry breaking of the Higgs mechanism. If this scale is much larger than the EW scale, the smallness of light neutrino masses would naturally stem from the suppression of the Majorana mass within the type-I Seesaw model~\cite{Minkowski:1977sc,Mohapatra:1979ia, Yanagida:1979as,Gell-Mann:1979vob}. Alternatively, in low-scale Seesaw realizations; such as the Inverse Seesaw~\cite{Mohapatra:1986aw,Mohapatra:1986bd,Bernabeu:1987gr} or the Linear Seesaw~\cite{Malinsky:2005bi}, with right-handed neutrinos below the EW scale, the smallness of light neutrino masses is explained through an approximate lepton number symmetry~\cite{Branco:1988ex,Kersten:2007vk,Abada:2007ux}. \\
The main phenomenological consequence of lowering the masses of these heavy neutral leptons (HNLs) is that they could be kinematically accessible to experiments, and can thus be produced and searched for. For a recent review of the present experimental status of HNLs see Ref.~\cite{Abdullahi:2022jlv}. \\
Moreover, HNLs could hold the key to understanding some other open problems of the SM and anomalies found in different experiments. These include, for instance, serving as candidates for dark matter~\cite{Abazajian:2012ys,denHerder:2009sxr}, or being able to account for the observed BAU via leptogenesis~\cite{Fukugita:1986hr}, even in its low-scale realizations~\cite{Akhmedov:1998qx,Asaka:2005pn,Shaposhnikov:2008pf}. Also, some anomalies recorded at short-baseline neutrino experiments, such as the Liquid Scintillator Neutrino Detector (LSND)~\cite{LSND:2001aii} and MiniBooNE~\cite{MiniBooNE:2008yuf}, have led to the proposal of HNLs, which could have a mass scale of the order of eV and participate in oscillations. However, introducing a single HNL to explain both the LSND and MiniBooNE anomalies is inconsistent with measurements of muon neutrino disappearance.\\
The existence of HNLs and the possibility of producing them in masses ranging from $\mathcal{O}(\text{MeV})$ to $\mathcal{O}(\text{TeV})$ is theoretically well motivated. Many experimental searches for HNLs have been carried out (a complete list of references with the most constraining results will be given in Section~\ref{sec:HNL_pheno}), but none have yielded evidence so far.
Instead, null results for searches have provided upper limits on the mixing of the HNLs with the active neutrinos as a function of their mass.

In this paper, we investigate the decays of HNLs into all possible final states within a liquid-argon time projection chamber (LArTPC)~\cite{Rubbia:1977zz} detector, such as the Imaging Cosmic And Rare Underground Signals (ICARUS) detector at the Short-Baseline Neutrino (SBN) Program~\cite{Rubbia:2015ihh} for the following reasons:\\
First:  The proposed physics objectives of the SBN Program offer exciting opportunities, including the capability to address certain experimental anomalies in neutrino physics, in particular with the ICARUS-T600 detector serving as the far detector of the SBN program. The ICARUS detector~\cite{ICARUS:2023gpo} is located off-axis to the high energy NuMI beam, and about 800~m away from the target. Although the primary focus of the ICARUS detector's physics program is to conduct the most sensitive search to date for sterile neutrinos at the eV mass-scale, the unique position of the ICARUS detector presents an excellent opportunity to explore other physics scenarios beyond the Standard Model, such as HNLs. To the best of our knowledge, this paper is  the first dedicated study to sensitivity for HNL searches at an ICARUS-like detector using the NuMI beam. \\
Second: The ICARUS detector is currently already collecting data at FNAL using both the BNB and the NuMI beams. We have considered NuMI beam for this study due to the following reasons: the NuMI beam has a higher average energy compared to the BNB, resulting in a greater mesons rate and potentially increased HNL production and the potential to study the tau mixing. Moroever, since the NuMI beam is off-axis relative to the ICARUS detector, the neutrino background will be significantly lower than that of the on-axis BNB beam. We have extensively studied all the possible decay channels of the HNLs and show the most promising channels covering a wide mass range $(10,  2000)$~MeV of the HNL masses, which can be analyzed using the presently collected ICARUS data.\\
Third: The ICARUS detector is the largest LArTPC detector operational in a neutrino beam  to date.  Using the pioneering LArTPC technology, with its full 3D imaging, excellent particle identification (PID) capability, and precise calorimetric energy reconstruction, the detector capabilities are well tailored
to identify all promising decay channels. Therefore, it is  very timely to initiate a search for HNLs using the available data at ICARUS.

This paper is organized as follows. In Section~\ref{sec:HNL_pheno}, the phenomenology of the HNL and the weak interactions that the HNL will inherit from the left-handed neutrinos via mixing are introduced, reviewing the present status of laboratory searches. In Section~\ref{sec:setup}, the experimental setup is introduced, explaining the relation between the coordinates of the ICARUS detector and those of the NuMI target. In Section~\ref{sec:signal_background}, we address the steps followed to simulate the signal and background events. The results are presented and discussed in Section~\ref{sec:results}. Finally, in Section~\ref{sec:conclusions}, we summarize the results and draw our conclusions.

\section{Benchmark HNL model}  
\label{sec:HNL_pheno}

Since neutrinos are the only neutral fermions of the SM, they are the only particles that could have either Dirac or Majorana masses. However; contrary to the rest of fermions that get their masses via Yuakawa interactions when the Higgs develops a vacuum expectation value $v$, a Dirac mass term of the form $-m_D\overline{\nu_L} \nu_R$ is not allowed for the neutrinos since the $\nu_R$ is not part of the SM particle content. On the other hand, a Majorana mass term for the left-handed neutrinos of the form $-\hat m \overline{\nu_L^c} \nu_L$, with $\nu_L^c = i\gamma^0\gamma^2 \overline{\nu_L}^t$, is not allowed in the SM Lagrangian by gauge invariance. Thus, neutrinos remain strictly massless in the SM, and we must go beyond the SM (BSM) to explain the overwhelming evidence for neutrinos masses and mixings observed in neutrino oscillation experiments.
The simplest of such extensions is to add heavy right-handed neutrinos, also known as HNLs, to the SM particle content within the type-I Seesaw model. In this work we will focus on the phenomenology consequences of extending the SM by one HNL. Notice that, even though at least two HNLs are required to explain the two mass splittings observed in neutrino oscillations, in this work we consider scenarios where the additional HNLs are too heavy to have an impact in low-energy observables. \\The flavor states will thus correspond to a combination of the light states and the additional heavy state
\begin{equation}
    \nu_{\alpha} = \sum_{i=1}^3 U_{\alpha i} \nu_{i} + U_{\alpha N} N\,,
    \label{eq:mixing}
\end{equation}
where $\alpha =e, \mu, \tau$ is the flavor index of the active neutrinos, $i$ is the mass index of the massive neutrinos, and $N$ is the HNL. The element of the neutrino mixing matrix $U_{\alpha N}$ parametrizes the mixing between the HNL and the active neutrino $\nu_\alpha$. The leptonic part of the Electroweak (EW) Lagrangian involving HNL interactions reads
\begin{equation}   
    \mathcal{L}_\text{EW} \supset - \frac{m_{W}}{v} U^{*}_{\alpha N} \overline{N} \gamma^{\mu} P_L \ell_{\alpha} W_{\mu}^+ - \frac{m_{Z}}{\sqrt{2}v} U^{*}_{\alpha N} \overline{N} \gamma^{\mu} P_L \nu_{\alpha} Z_{\mu} \,,
    \label{eq:interaction}
\end{equation}
with $\ell_\alpha$ the charged leptons, $P_L$ the left projector, and where $m_W$ and $m_Z$, are the $W$ and $Z$ gauge boson masses, respectively. Eq.~(\ref{eq:interaction}) shows that the HNL will participate in charged current (CC) and neutral current (NC) interactions in which the active neutrinos appear, with an additional suppression from the mixing $U_{\alpha N}$. These elements are typically small, giving rise to weaker-than-weak interactions, suppressed production, and long lifetimes due to its suppressed decay width. Nevertheless, laboratory experiments can be sensitive to an HNL in a wide mass range. When the mass of the HNL is very light ($\mn \lesssim 10$~eV), it participates in neutrino oscillations modifying the standard $3\nu$-oscillation picture (see for instance the results~\cite{LSND:2001aii,NOMAD:2001xxt,KARMEN:2002zcm,NOMAD:2003mqg,MiniBooNE:2007uho,Tsenov:2009jca,IceCube:2016rnb,MINOS:2020iqj,MiniBooNE:2020pnu}). For heavier HNLs, the oscillation is too fast to be resolved at the detector, and the bound on the mixing becomes independent of its mass (averaged-out regime). For long-baseline neutrino oscillation experiments, this regime typically occurs for new squared-mass differences between the HNL and the light neutrinos of $\Delta m^2\gtrsim \mathcal{O}(100)~\text{eV}$. Even though this bound always applies for heavier HNLs, we enter the regime where other observables set stronger constraints.
 For HNL masses up to MeV-scale, it would induce a kink in the Kurie plot at the end point of the electron energy spectrum of nuclear $\beta$-decay (see for instance the review in Ref.~\cite{Atre:2009rg} and references therein). For HNL masses up to the GeV-scale, they could be produced in meson decays. In particular, they would modify the energy spectrum of two-body decays of pseudoscalar mesons, producing additional peaks in the spectrum (see for instance the following experimental results~\cite{Daum:1987bg,Britton:1992xv,E949:2014gsn,PIENU:2017wbj,PIENU:2019usb,NA62:2020mcv}). Also, when the HNL originates from the decay of mesons produced in a beam-dump experiment, since the HNL is quite energetic, its boosted decay length could range from a few meters up to 1~km, opening the possibility to decay in-flight in the near detector (ND) of a neutrino experiment (see for instance~\cite{WA66:1985mfx,NuTeV:1999kej}). Notice however, that the number of decay events for in-flight searches is proportional to $\vert U_{\alpha N}\vert^4$, since the HNL has to be produced and decay back into SM particles. Colliders look for up to TeV-scale HNLs produced via $W$ or $Z$ interactions, and decaying back to SM producing signatures with displaced vertices (see for instance~\cite{ATLAS:2022atq,CMS:2024bni}) or, if Majoranan HNL, same sign lepton pairs (see for instance~\cite{CMS:2024bni,ATLAS:2024erm}). Finally, when $\mn > \text{EW}$ scale, they would produce deviations on EW and flavor precision observables through the non-Unitarity of the PMNS mixing matrix~\cite{Broncano:2002rw,Broncano:2003fq,Antusch:2006vwa,Fernandez-Martinez:2007iaa,Abada:2007ux}. \\
Experimental collaborations typically assume so far single flavor dominance when providing limits on a particular mixing matrix element $\ua2$. That is, the mixing with the flavor $\alpha$ is assumed to dominate over the other two flavors. Even though, these simplified flavor structure benchmarks are in tension with present neutrino oscillation data in minimal neutrino mass models~\cite{Drewes:2022akb}, we have chosen to show our results following this simplified scenario for easier comparison with the present experimental results. Thus, the benchmark HNL model assumed in this paper has four free parameters, the three mixing matrix elements $\ua2$, and the mass of the HNL $\mn$. Also in this work the HNL is assumed to be Dirac unless otherwise specified. Since in the case of Majoranan HNL, the decay width gets an additional factor of two, the results obtained in this work can be extrapolated to the Majorana case simply by re-scaling the expected sensitivity of Dirac HNL by a factor $1/\sqrt{2}$.\\
In the following, we provide a list of the present experiments setting the most stringent constrains on $\ua2$ as a function of $\mn$ in the range $\sim \mathcal{O}(\text{MeV},\text{TeV})$:

\begin{itemize}
    \item $\ue2$ : $\pi$ and $K$  universality tests (Bryman-Shrock)~\cite{Bryman:2019bjg}, 
    ATLAS (2019)~\cite{ATLAS:2019kpx}, 
    ATLAS (2022)~\cite{ATLAS:2022atq}, 
    BEBC (Barouki et al)~\cite{Barouki:2022bkt}, 
    Belle~\cite{Belle:2013ytx}, 
    Borexino~\cite{Borexino:2013bot}, 
    CHARM~\cite{CHARM:1985nku}, 
    CMS (2018)~\cite{CMS:2018iaf}, 
    CMS (2022)~\cite{CMS:2022fut}, 
    CMS (2024-I)~\cite{CMS:2024ake}, 
    CMS (2024-II)~\cite{CMS:2024xdq}, 
    DELPHI~\cite{DELPHI:1996qcc}, 
    L3 (2001)~\cite{L3:2001zfe}, 
    LSND (Ema et al)~\cite{Ema:2023buz}, 
    NA62~\cite{NA62:2020mcv}, 
    PIENU (2017)~\cite{PIENU:2017wbj}, 
    PMNS Unitarity (Blennow et al)~\cite{Blennow:2023mqx}, 
    super-allowed $\beta$ decays (Bryman-Shrock)~\cite{Bryman:2019bjg}, 
    T2K~\cite{T2K:2019jwa}, 
    TRIUMF~\cite{Britton:1992xv}.
    
    \item $\um2$: $\mu$BooNE~\cite{MicroBooNE:2023eef}, 
    ATLAS (2019)~\cite{ATLAS:2019kpx}, 
    ATLAS (2022)~\cite{ATLAS:2022atq}, 
    BEBC~\cite{WA66:1985mfx}, BNL-E949~\cite{E949:2014gsn}, 
    CHARM-II~\cite{CHARMII:1994jjr}, 
    CMS (2018)~\cite{CMS:2018iaf}, 
    CMS (2018-dilepton)~\cite{CMS:2018jxx}, 
    CMS (2022)~\cite{CMS:2022fut}, 
    CMS (2024-I)~\cite{CMS:2024ake}, 
    CMS (2024-II)~\cite{CMS:2024xdq}, 
    CMS (8~TeV)~\cite{CMS:2016aro}, 
    DELPHI (short)~\cite{DELPHI:1996qcc},
    KEK~\cite{Bryman:2019bjg}, 
    LSND (Ema et al)~\cite{Ema:2023buz}, 
    NA3~\cite{NA3:1986ahv}, 
    NA62~\cite{NA62:2021bji}, 
    NuTeV~\cite{NuTeV:1999kej}, PIENU~\cite{PIENU:2019usb}, 
    PMNS Unitarity (Blennow et al)~\cite{Blennow:2023mqx}, PSI~\cite{Daum:1987bg},
    T2K~\cite{T2K:2019jwa}, 
    T2K (Arg\"uelles et al)~\cite{Arguelles:2021dqn}.

    \item  $\ut2$: ArgoNeuT~\cite{ArgoNeuT:2021clc}, 
    Atmospheric $\nu$ (Dentler et al)~\cite{Dentler:2018sju}, 
    BEBC (Barouki et al)~\cite{Barouki:2022bkt}, 
    BaBar ~\cite{BaBar:2022cqj}, Belle~\cite{Belle:2024wyk}, 
    Borexino (Plestid)~\cite{Plestid:2020ssy}, 
    CHARM (Boiarska et al)~\cite{Boiarska:2021yho}, 
    CHARM (Orloff et al)~\cite{Orloff:2002de}, 
    DELPHI~\cite{DELPHI:1996qcc} 
    PMNS Unitarity (Blennow et al)~\cite{Blennow:2023mqx}.
\end{itemize}

\section{Experimental setup}
\label{sec:setup}

The LArTPC technology has become widely used in modern accelerator neutrino experiments due to its excellent calorimetric reconstruction and spatial resolution. The current and near-future Fermilab neutrino program uses LArTPCs for neutrino oscillation measurements at both short and long neutrino baselines. These detectors are also being deployed in exotic signature searches, allowing for the exploration of previously untested parameter space of BSM scenarios.\\ 
The ICARUS detector is the third detector; together with the detectors from the Micro-Booster Neutrino Experiment (MicroBooNE) and Short-Baseline Near Detector (SBND) experiment, exposed to the Booster Neutrino Beam (BNB) $\nu$-beam of the Short-Baseline Neutrino program at Fermilab. The SBN was proposed to investigate the possible existence of sterile neutrinos in the mass region of around 1~eV, and to clarify the LSND and MiniBooNE anomalies. ICARUS is the largest detector of the SBN Program, a rectangular box with internal dimensions of $(x,y,z) = (3.6, 3.9, 19.6)$~m, filled with 760 tons of liquid argon. The size and fiducial mass of ICARUS makes it interesting for long-lived particles (LLPs) searches~\cite{Berger:2024xqk,Batell:2019nwo}.\\
In this paper, we study the capabilities of an ICARUS-like detector to look for HNLs produced in the neutrinos at the Main Injector (NuMI) beam~\cite{Adamson:2015dkw}. While BNB utilizes a proton energy of 8~GeV, NuMI produces a high energetic $\nu$-beam from the collision of the 120~GeV protons with a graphite target. The produced hadrons are focused by two magnetic horns before entering a 675~m long decay volume. The horn system selects the charge of the mesons produced in the target, with running while focusing positively-charged mesons referred to as “neutrino mode”, and running while focusing negatively-charged mesons referred to as “antineutrino mode”. At the end of the decay pipe there is 5~m thick absorber to attenuate the residual hadrons. In order to point to the far detector (FD) of MINOS at the Soudan Underground Laboratory in Minnesota, the NuMI beam is inclined downward by 58~mrad. \\
In this work, the origin of the ICARUS detector coordinates system are fixed at the center of the detector, with $(x_I,y_I,z_I) = (\text{west}, \text{up}, \text{north: along BNB axis})$. Instead, the NuMI coordinates, $(x_N,y_N,z_N)$, are centered at the NuMI target, with $z_N$ along the NuMI beam axis (northwest, downward 58~mrad). The two sets of coordinates are related by the following transformation~\footnote{Internal discussion with NuMI beam group.}

\begin{eqnarray}
\left(\begin{array}{c} x_I \\ y_I \\ z_I \end{array} \right) &=&
    \begin{pmatrix}
        0.92 & 0.02 & 0.39\\
        0 & 1 & -0.06 \\
        -0.38 & 0.05 & 0.92
    \end{pmatrix} \left(\begin{array}{c} x_N \\ y_N \\ z_N \end{array} \right) \nonumber \\
  & &- \left(\begin{array}{c} 315.12 \\ 33.64 \\ 733.63 \end{array} \right) [m]\,.
    \label{eq:coordinates}
\end{eqnarray}

\begin{center}
    \begin{figure*}[htb!]
        \includegraphics[width=0.8\textwidth]
        {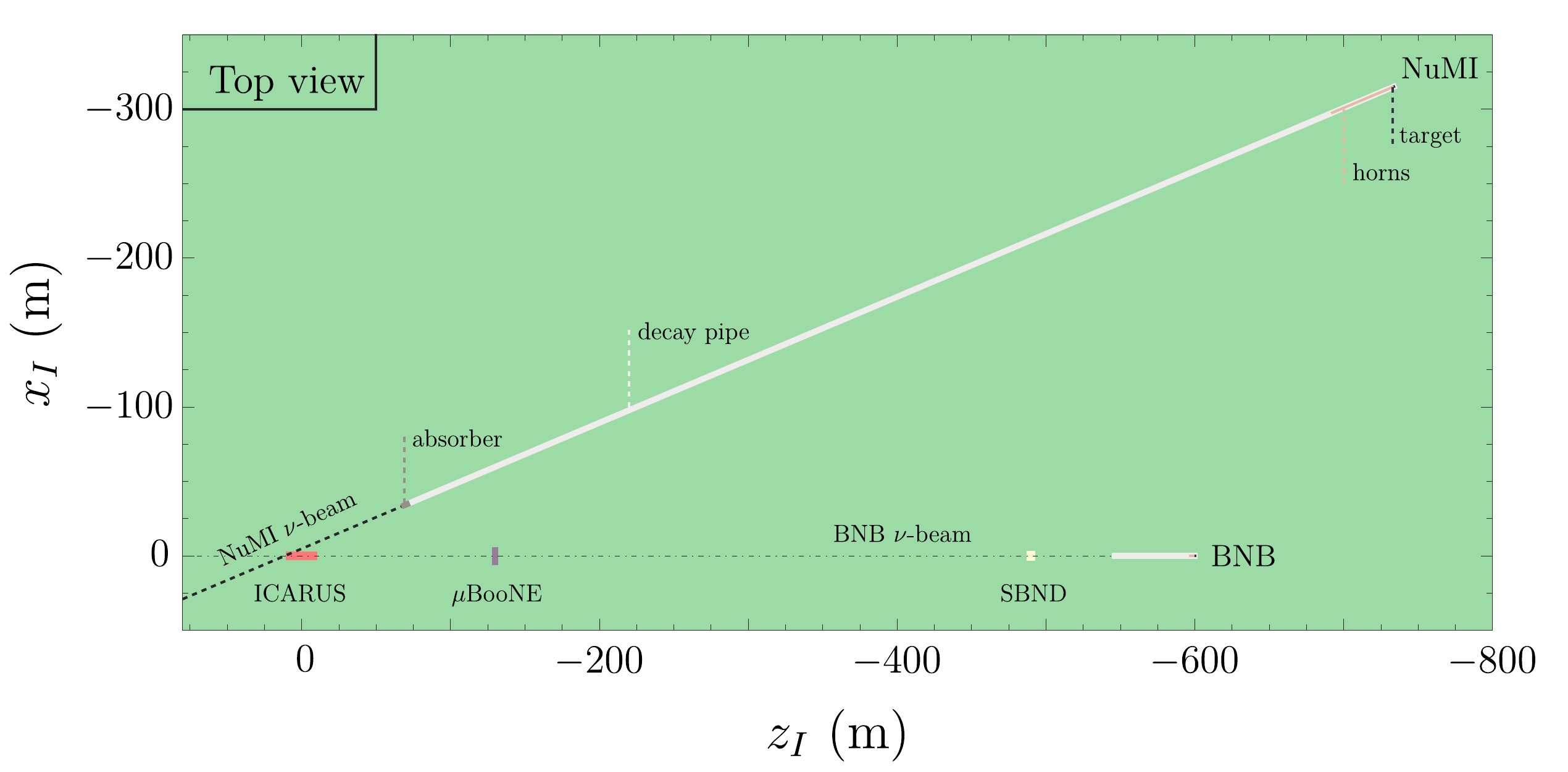}
        \includegraphics[width=0.8\textwidth]
        {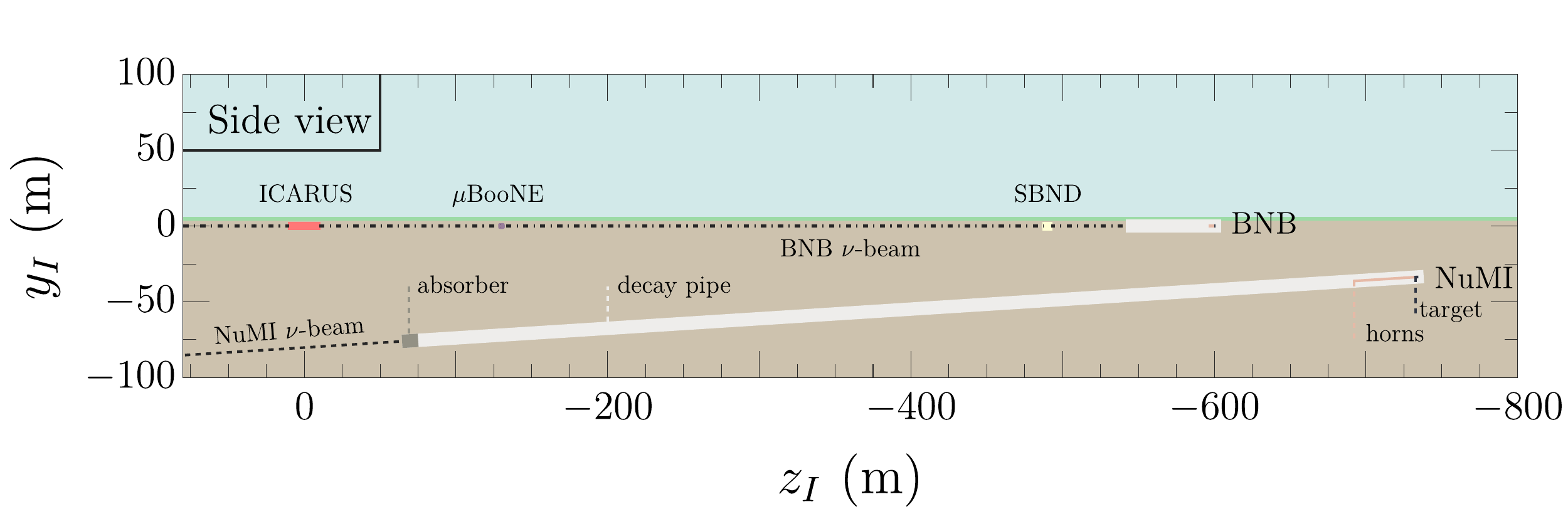}
    \caption{Sketch of the experimental configuration considered in this work. The top view (upper panel), and side view (lower panel) correspond to the $x_Iz_I$ and $y_Iz_I$ projections of the ICARUS set of coordinates, respectively. Drawn to scale.}
    \label{fig:setup}
    \end{figure*}
\end{center}

 The sketch of the experimental setup considered in this study is shown in Figure~\ref{fig:setup}. The upper (lower) panel represents the top (side) view of the Fermilab area where the BNB and NuMI beam lines are placed, as seen from the $x_I z_I$ ($y_I z_I$) projection of the ICARUS set of coordinates. All the elements and distances in Figure~\ref{fig:setup} have been drawn to scale. \\
 The SBN Program proposal states~\cite{Rubbia:2015ihh} that ICARUS will receive $6.6 \cdot 10^{20}$ protons on target (PoT) per year from NuMI neutrino beam. In this paper, we present a search for HNL using the NuMI beam data corresponding to $1.32\cdot 10^{21}$ PoT with the ICARUS  detector. This means that our result would correspond to 2 years of ICARUS data collection.

\section{Simulations}
\label{sec:signal_background}

Signal and background processes have been simulated, as discussed in this section, and a simple event selection has been used to evaluate the signal efficiency and background rejection in the ICARUS-like detector for the HNL decay channels. This requires the simulation of the fluxes of HNL arriving into the ICARUS-like detector, and the distributions of their corresponding decay products, as well as the simulation of the main potential background that could mimic the HNL signals in the detector. 

\subsection{Signal events}
\label{subsec:signal}

The first step in simulating the signal events is to simulate the fluxes of particles produced at the NuMI target after the collision, and which will decay into neutrinos. For pions and kaons, we utilize the publicly available 
\textsc{Geant4}-based~\cite{GEANT4:2002zbu,Allison:2006ve,Allison:2016lfl} package prepared by the NuMI beam department; G4NuMI~\cite{AliagaSoplin:2016shs}, which simulates the interaction of 120~GeV protons on the graphite target, the production of mesons, and the re-interaction and focusing of those mesons in the magnetic horns. This pool of events contains the coordinates where the mesons decay into neutrinos, and their corresponding momenta. 
On the contrary, the heavier parents, $D$ and $D_s$; and the $\tau$ leptons produced from their decay, are not included in  G4NuMI. Thus, we have adopted the simulations of heavy parents at LBNF performed for Ref.~\cite{Coloma:2020lgy}. In this reference, the fluxes of heavy parents were obtained by simulating the inelastic interactions of the protons in the target with \textsc{Geant4}, and then passing them to \textsc{Pythia8.2}~\cite{Sjostrand:2014zea} to compute the production rate of the parents. This simulations also include the additional heavy mesons produced from the collision of the remaining protons passing through the target and arriving into the absorber, located at the end of the decay pipe. In this analysis, we have adapted and normalized the simulations to the correct luminosity, target size, and geometry of the decay pipe and absorber of NuMI. Table~\ref{tab:parent_yield} collects the averaged production yield per PoT of the different parents that would produce neutrinos towards the region of ICARUS, during the neutrino mode of NuMI.

\begin{table}[htb!]
\begin{center}
\begin{small}
\renewcommand{\arraystretch}{1.8}
\begin{tabular}[t]{ | c || c | c | c | c | c |}
\hline
 & $\pi$ & $K$ & $D$ & $D_s$ & $\tau$ \\
 \hline
 \hline
$P^+/$PoT & 6.0 & 1.1 & $1.2\cdot10^{-5}$ & $3.3\cdot10^{-6}$ & $2.1\cdot10^{-7}$\\
\hline
$P^-/$PoT & 4.1 & 0.4 & $1.9\cdot10^{-5}$ & $4.6\cdot10^{-6}$ & $3.0\cdot10^{-7}$\\
\hline
\end{tabular}
\caption{Averaged parent production yield at the NuMI target per PoT during its neutrino mode. A pre-selection of light mesons that would produce neutrinos towards the region where ICARUS is placed has been done.}
\label{tab:parent_yield}
\end{small}
\end{center}
\vspace{-0.4cm}
\end{table}

The second step is to utilize the FeynRules~\cite{Alloul:2013bka} model file within \texttt{HNLux}, a tool developed for Ref.~\cite{Coloma:2020lgy} to simulate the HNL fluxes. \texttt{HNLux} was used to generate \textsc{MadGraph5}~\cite{Alwall:2014hca} events for $N$ production and $N$ decay. For each mixing matrix element $\ua2$, we generated a grid of more than 200 mass points within the mass window $\mn\in(10,2000)$~MeV; and for each mass, $10^6$ \textsc{MadGraph5} events were simulated. The HNL production channels included in our simulations are listed in Table~\ref{tab:HNL_production}. These correspond to the leptonic and semileptonic decays of the parent mesons, and the tau leptons with higher branching ratio to neutrinos.  The decay channel $\tau^- \to \pi^- \pi^0 N$ is simulated via the approximation $\tau^-\to \rho^-  N\,, ~\rho^- \to \pi^- \pi^0$, as discussed in~\cite{Coloma:2020lgy}. 

\begin{table}[htb!]
\begin{center}
\begin{small}
\renewcommand{\arraystretch}{1.8}
\begin{tabular}[t]{ | c || c c c c c | }
\hline

 & $\pi\to$ & $K\to$ & $D\to$ & $D_s\to$ & $\tau\to$ \\
 \hline
 \hline
\multirow{2}{*}{$\ue2$} & $e N$ & $e N$ & $e N$ & $e N$ & \textemdash \\
 & \textemdash & $\pi^0 e N$ & $K^0 e N$ & \textemdash & \textemdash \\
\hline
\multirow{2}{*}{$\um2$} & $\mu N$ & $\mu N$ & $\mu N$ & $\mu N$ & \textemdash \\
 & \textemdash & $\pi^0 \mu N$ & $K^0 \mu N$ & \textemdash & \textemdash \\
\hline
\multirow{4}{*}{$\ut2$} & \textemdash & \textemdash & $\tau N$ & $\tau N$ & $\pi N$ \\
 & \textemdash & \textemdash & \textemdash & \textemdash & $\pi\pi^0 N$ \\
 & \textemdash & \textemdash & \textemdash & \textemdash & $e \nu N$ \\
 & \textemdash & \textemdash & \textemdash & \textemdash & $\mu \nu N$ \\
\hline
\end{tabular}
\caption{List of 2-body and 3-body decays of the parents into an HNL considered in this work.}
\label{tab:HNL_production}
\end{small}
\end{center}
\vspace{-0.4cm}
\end{table}

On the other hand, the HNL decay channels with $\mathcal{B}_N\gtrsim 1\%$ considered in this work are summarized in Table~\ref{tab:HNL_decays}. Notice that; even though $N\to \nu\nu\nu$ almost dominate the entire HNL mass region studied in this work, it has not been included in the simulations since it would not produce visible signatures at the ICARUS detector. The HNL decay channels $N\to \ell^\pm \text{hadr.}$ or $N\to \nu \text{hadr.}$ corresponds to $N$ charged or neutral current decays, with three or more mesons in the final state, respectively. These channels become relevant for $\mn\gtrsim 1.3$~GeV.  

\begin{table}[htb!]
\begin{center}
\begin{small}
\renewcommand{\arraystretch}{1.8}
\begin{tabular}[t]{ | c || c  c  c  c  c | }
\hline
 &  & & $N\to$ &  &  \\
 \hline
 \hline
\multirow{2}{*}{$\ue2$} & $e e \nu$ & $\mu \mu \nu$ & $e \mu \nu$ & $e~\text{hadr.}$  &  $\nu~\text{hadr.}$ \\
 & $\nu \pi^0$ & $e \pi$ & $e \rho$ & \textemdash &  \\
\hline
\multirow{2}{*}{$\um2$} & $e e \nu$ & $\mu \mu \nu$ & $e \mu \nu$ & $\mu~\text{hadr.}$  &  $\nu~\text{hadr.}$ \\
 & $\nu \pi^0$ & $\mu \pi$ & $\mu \rho$ & \textemdash &  \\
\hline
\multirow{2}{*}{$\ut2$} & $e e \nu$ & $\mu \mu \nu$ & \textemdash & \textemdash  & $\nu~\text{hadr.}$ \\
 & $\nu \pi^0$ & \textemdash & \textemdash & $\nu \rho^0$ &   \\
\hline
\end{tabular}
\caption{List of the most relevant $N$ decays into SM particles.}
\label{tab:HNL_decays}
\end{small}
\end{center}
\vspace{-0.4cm}
\end{table}

\texttt{HNLux} operates as follows: for each event of parent meson produced at the target, a random \textsc{MadGraph5} event of $N$ production is selected. The position and momentum of the meson are then used to compute the momentum of the HNL boosted to the lab frame, which will define the trajectory followed by the HNL. If this trajectory intersects the geometry of the ICARUS detector, the event will be saved, creating a pool of HNL crossing the detector. For each of these events, a random \textsc{MadGraph5} event of $N$ decay is then selected. The momentum of the HNL decay products are then computed in the lab frame. Also a random point inside the trajectory crossing the detector is selected, which defines the coordinates where the HNL decays into the corresponding decay products of the channel considered in the simulations. 

\subsection{Background events}
\label{subsec:background}

The primary background source in this search will be CC and NC interactions of SM neutrinos from the NuMI $\nu$-beam with the $^{40}\text{Ar}$ nuclei within the active volume of the ICARUS TPCs. Other potential background sources; such as neutrino interactions with rock or cosmic muons, are expected to be negligible in comparison, as the resulting events will generally not align with the direction of the beam. The \textsc{Genie}~\cite{Andreopoulos:2009rq} Monte Carlo neutrino event generator and the public NuMI light neutrino flux files have been used to simulate events of $\nu_\alpha$-$^{40}\text{Ar}$ interactions. At these energies the main processes to be considered are:
\begin{itemize}
    \item Single charged pion production from $\nu_\alpha$ resonant or coherent scatterings with $^{40}\text{Ar}$ nuclei.  The signature of this process is $\pi^+ \ell_\alpha^-$, where $\alpha$ denotes the flavor of the incoming active neutrino.
    \item Single neutral pion production from $\nu_\alpha$ NC resonant scatterings with $^{40}\text{Ar}$ nuclei.
    \item Charm production in CC deep inelastic $\nu_\alpha$-$^{40}\text{Ar}$ nucleus scattering (DIS). The subsequent prompt decay of the charm meson will produce an additional charged lepton in the final state. Thus, the signature of this process would be $\ell^-_\alpha\ell^+_\beta$, with $\alpha$ the flavor of the incoming SM neutrino, and $\beta$ the flavor of the secondary charged lepton produced via charm meson decay. Notice however that this process is suppressed in comparison with the single $\pi^+$ or $\pi^0$ production processes.  
\end{itemize}

The convolution of the active neutrino flux crossing the fiducial volume of ICARUS, along with the CC and NC $\nu_\alpha$-$^{40}\text{Ar}$ cross sections, provides the expected number of neutrino interactions with the liquid argon in the TPCs. Table~\ref{tab:nu_interactions} summarizes the expected number of $\nu_e$-$^{40}\text{Ar}$ and $\nu_\mu$-$^{40}\text{Ar}$ CC and NC interactions per year ($\nu_\tau$ contribution is negligible). Wrong-sign backgrounds from misidentified events initiated by the $\overline{\nu_{e}}$ and $\overline{\nu_{\mu}}$  beam components in neutrino mode have been included  in the simulations. 

\begin{table}[htb!]
\begin{center}
\begin{small}
\renewcommand{\arraystretch}{1.8}
\begin{tabular}[t]{ | c || c | c |}
\hline
 & CC $\nu_\alpha$-$^{40}\text{Ar}$  & NC $\nu_\alpha$-$^{40}\text{Ar}$ \\
 \hline
 \hline
$\alpha = e$ & 19142 & 2109 \\
\hline
$\alpha = \mu$ & $341257$ & 42610 \\
\hline
\end{tabular}
\caption{Expected number of $\nu_\alpha$-$^{40}\text{Ar}$ CC and NC interactions/year at ICARUS detector from NuMI beam, with $\alpha=e~\text{or}~\mu$.}
\label{tab:nu_interactions}
\end{small}
\end{center}
\vspace{-0.4cm}
\end{table}

\section{Results}
\label{sec:results}

The total expected number of HNL decays into a given decay channel $c$ inside the ICARUS detector can be expressed as
\begin{equation}
    N^c =\mathcal{B}_N^c \times \int dE_N P(E_N) \frac{d\phi_N}{dE_N} \, ,
    \label{eq:N_events}
\end{equation}
where $\mathcal{B}_N^c$ is the branching ratio of the corresponding decay channel $c$, and $P(E_N)$ is the probability of the HNL to decay inside the volume of the detector given by
\begin{equation}
    P(E_N) = e^{-\frac{\Gamma L}{\gamma \beta}}\left(1-e^{-\frac{\Gamma \Delta \ell_\text{det}}{\gamma \beta}}\right)\,,
    \label{eq:probability}
\end{equation}
with $\Gamma$ the total decay width of the HNL in its rest frame, $L$ the distance traveled by the HNL between its production and its intersection with ICARUS detector, and $\Delta \ell_\text{det}$ the length of detector crossed by the HNL trajectory. The energy dependence in Eq.~(\ref{eq:probability}) enters through the boost factor $\gamma \beta = \sqrt{E_N^2/\mn^2-1}$.

To calculate the expected sensitivity to the $\ua2$ and the mass of the HNL, we conduct an unbinned Gaussian $\chi^{2}$ for each detection channel, defined as
\begin{equation}
    \chi^{2}= min_{\xi}\Biggl\{\Biggl(\frac{N^{ex}(\phi, \xi) - N^{ob}(\xi)}{\sigma_{N}(\xi)}\Biggl)^{2} + \Biggl(\frac{\xi}{\sigma_{f}}\Biggl)^{2}\Biggl\},
    \label{eq:chi2_bg}
\end{equation}
where $N^{ex}(\phi, \xi)$ is the total expected event rate including both signal and background events, and $N^{ob}(\xi)$ is the observed events; with the uncertainty on the flux normalization, $\sigma_{f}$, taken as $20\%$, and $\sigma_{N}(\xi) = \sqrt{N^{ob}(\xi)}$ is the statistical error. We take the observed events as an expectation in the absence of an HNL signal. Here $\phi$ corresponds to the model parameters ($\mn$ and $\ua2$), while $\xi$ is the nuisance parameter, which accounts for the systematic uncertainties affecting their overall normalization.  Our sensitivity regions are obtained taking the corresponding $\chi^{2}$ cut at a given confidence level (C.L.) with 2 degrees of freedom (d.o.f.). In particular, we set the 90\% C.L. contour by taking the contour at $\chi^{2}=4.61$. Finally, for the analysis where no background is considered, we follow the Feldman and Cousins prescription~\cite{Feldman:1997qc} for a Poisson distribution with no background and under the hypothesis of no events being observed, which corresponds to
\begin{equation}
    N^{ex}(\phi) < 2.44\,, \quad \text{for}~90\%~\text{C.L.}
    \label{eq:Feldman-Cousins}
\end{equation}

\subsection{The role of the HNL decay channel}
\label{subsec:sensitivity_channel}

\begin{center}
    \begin{figure*}[htb!]
        \includegraphics[width=0.84\textwidth]   {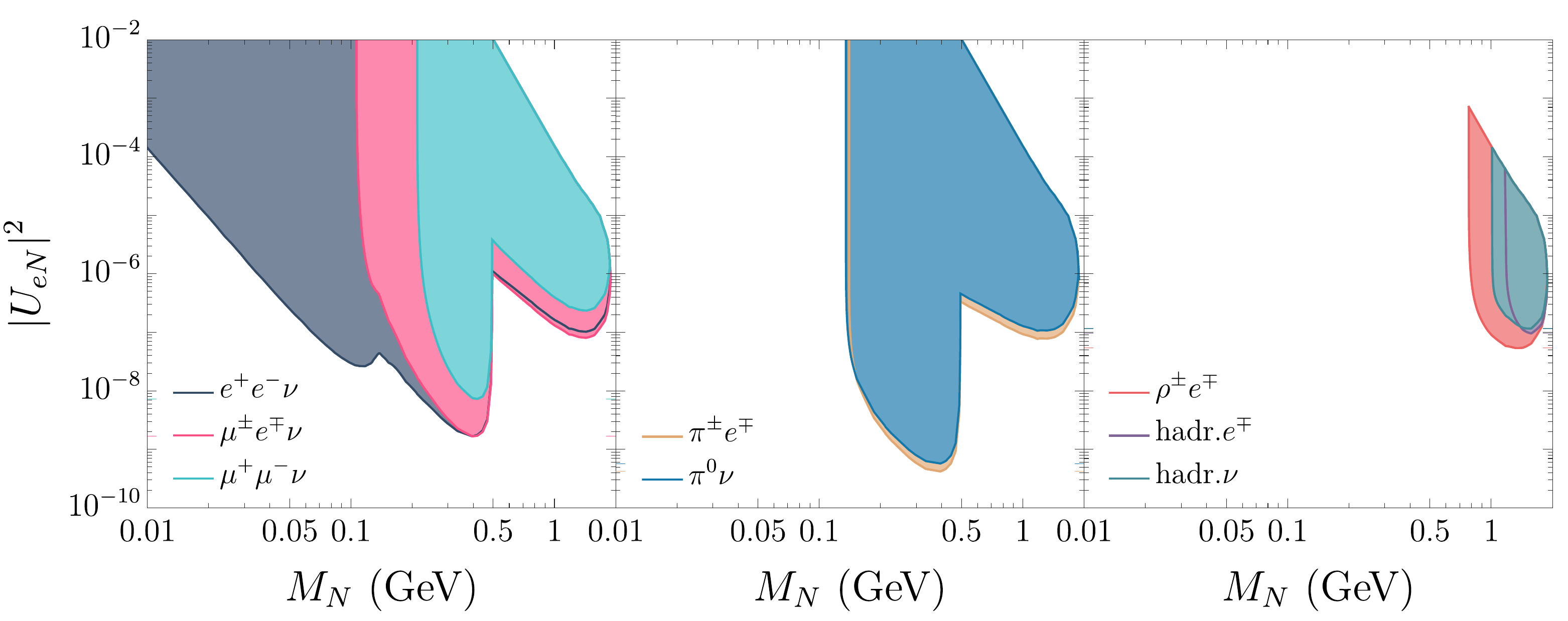}
        \includegraphics[width=0.84\textwidth]   {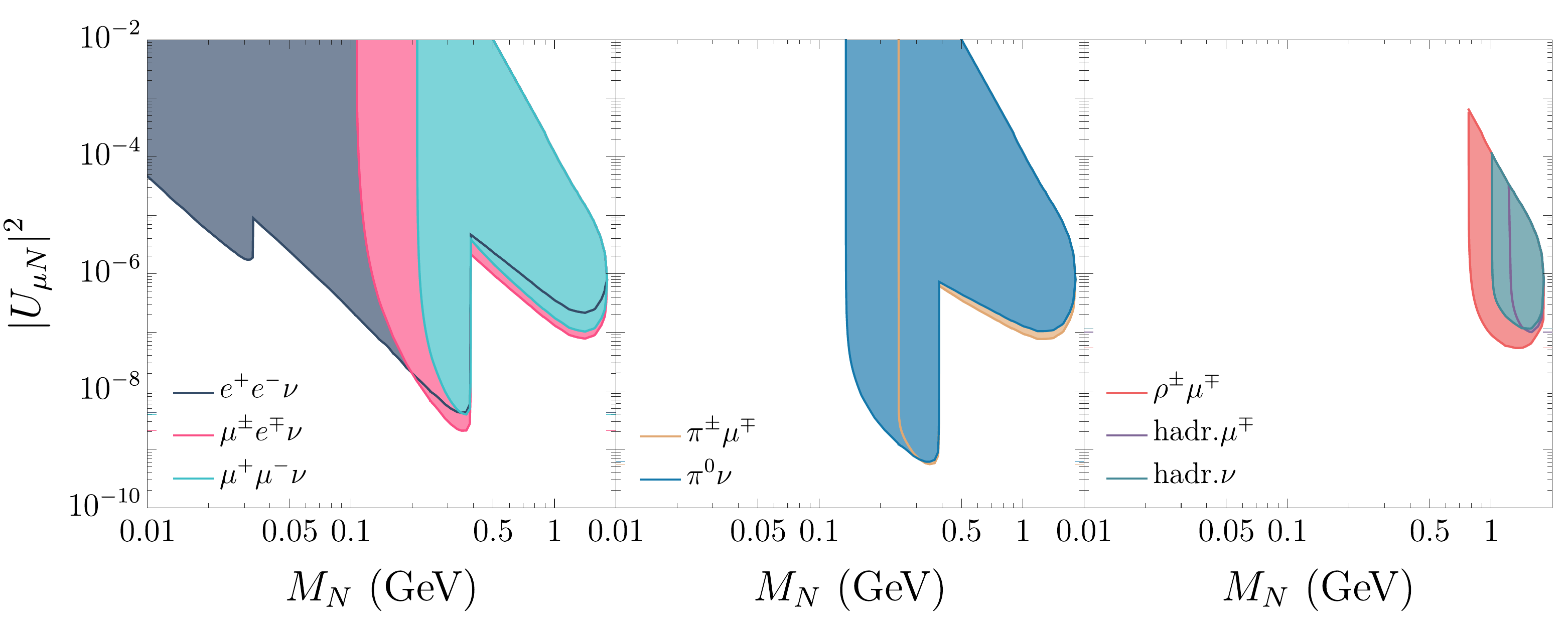}
        \includegraphics[width=0.84\textwidth]   {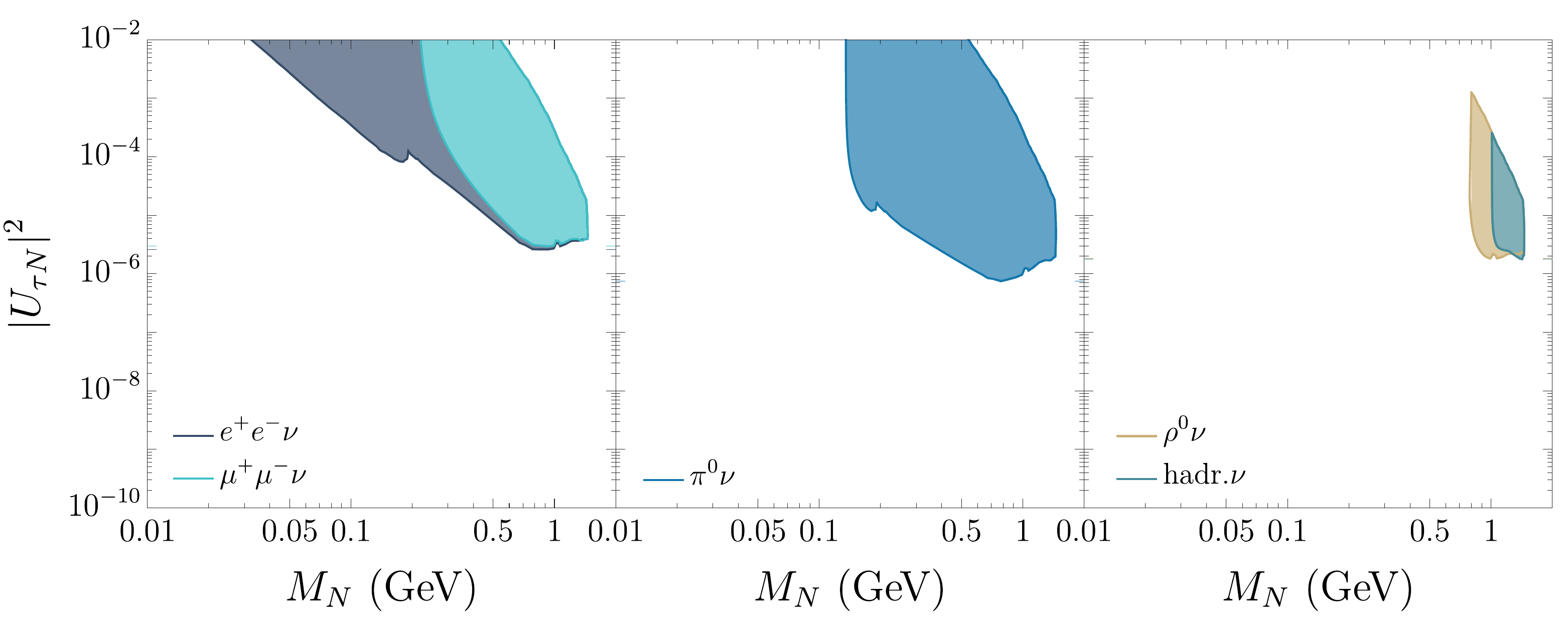}
        \caption{ICARUS-like sensitivity at 90$\%$ C.L. level for different mixing matrix elements as a function of $\mn$, based on $1.32 \cdot 10^{21}$ PoT from the NuMI beam. In each row, we assume that the HNL only couples to one of the charged leptons as indicated, while the other two mixings are set to zero. Left panels correspond to HNL decays with charged leptons and missing energy, while middle (right) panels correspond to HNL decays with pseudoscalar (vector) mesons in the final state.}
        \label{fig:signal_only}
    \end{figure*}
\end{center}

It is crucial to highlight the significance of the various final decay channels for the HNLs. Figure~\ref{fig:signal_only} shows the ICARUS expected sensitivity contours in the $\mn - \ua2$ plane at  $90\%$ C.L. for $1.32 \cdot 10^{21}$ protons on target (PoT) from the NuMI beam, assuming that the HNL mixes with $e$ (top), $\mu$ (middle) or $\tau$ (bottom),  respectively. Left panels correspond to HNL decays with charged leptons and missing
energy, while middle (right) panels correspond to HNL decays with pseudoscalar (vector) mesons in the final state. For a given flavor of the mixing, and value of $\mn$; the production of the HNL is the same for left and right panels, being the $\mathcal{B}_N^c$ in Eq.~(\ref{eq:N_events}) their only difference. Thus; for each $\mn$, the sensitivity will be dominated by the HNL decay channels with higher branching ratio. In this analysis, we have assumed exclusive mixing of a heavy neutrino with one active neutrino, setting the other two mixings to zero. Also, the contours have been obtained following Eq.~(\ref{eq:Feldman-Cousins}); that is, by counting the number of events without any background consideration.

The panels with $e$ and $\mu$ mixing show three different sensitivity regions 
\begin{itemize}
    \item Region I: $\mn< m_{\pi} - m_{\ell_\alpha}$; where the leading contribution to the sensitivity comes from $\pi^\pm$ decays due to its larger production yield (see Table~\ref{tab:parent_yield}). In this region, the only relevant detection channel for the HNL is the $N\to e^+e^-\nu$.
    \item Region II: $m_{\pi} - m_{\ell_\alpha} < \mn< m_{K} - m_{\ell_\alpha}$; where the phase space of the $\pi^\pm$ channel is closed, and the leading HNL production comes from $K^\pm$ decays. Notice that, even though the production yield of $K^\pm$ is smaller than that of $\pi^\pm$, the sensitivity reaches its best point (lower limit) due to the fact that HNLs of these masses are less long-lived, and thus they decay within the distance of ICARUS. In this region, the dominant HNL decay channels leading the sensitivity are: $N\to e^+e^- \nu$ (just for $\mu$ mixing), $N\to \pi^0 \nu$, and $N\to \pi^\pm \ell^\mp_\alpha$.
    \item Region III: $\mn > m_{K} - m_{\ell_\alpha}$; where the HNLs are produced via $D^\pm$ and $D_s^\pm$ meson decays. The lower production yield of heavy mesons at the NuMI target significantly reduces the expected sensitivity in this region. In this region, the sensitivity is first dominated by $N\to \pi^0 \nu$, and $N\to \pi^\pm \ell^\mp_\alpha$ decay channels; while for $\mn\geq 1$~GeV, the channels $N\to e^\mp \mu^\pm \nu$, $N\to  \ell^\mp_\alpha\ell^\pm_\alpha \nu$, and $N\to \rho^\pm \ell^\mp_\alpha$ join the leading contribution to the sensitivity. 
\end{itemize}

In contrast, the panel with $\tau$ mixing shows two different sensitivity regions 
\begin{itemize}
    \item Region I: $\mn< m_{D_s} - m_{\tau}$; where the production of HNL comes from $D^\pm$ and $D_s^\pm$ decays into $\tau^\pm$ lepton, and where the sensitivity is mostly dominated by the $N\to e^+e^- \nu$ HNL decay channel.
    \item Region I:  $m_{D_s} - m_{\tau} < \mn<  m_{\tau} - m_{e}$; where the production of HNL comes from $\tau^\pm$ decays, and where the sensitivity is mostly dominated by the HNL decay channel $N\to  \pi^0 \nu$.  
\end{itemize}

The sensitivity curve in other configurations of the mixing different from the single flavor dominance one could be estimated from the individual results shown in Figure~\ref{fig:signal_only}. Remember that for a given HNL decay channel into SM particles, the number of HNL decay events inside the detector volume is proportional to the product of the parent decay width into HNL, and the HNL decay width into the specific visible channel. And notice that the decay width of charged pseudoscalar mesons depends on the mass of the lepton in the final state. Thus, in the case of degenerate mixing ($\ue2 = \um2 = \ut2$), the parent decay will be dominated by the contribution of the heaviest charged meson allowed by phase space. Additionally, the HNL decay width of channels with neutrinos in the final state will be enhanced due to the extra contributions from the possible diagrams. Consequently, in the particular case of degenerate mixing, the sensitivity lines of the individual channels shown in Figure~\ref{fig:signal_only} will follow the minimum value among those of the electron, muon, and tau panels.
That is, in each region I/II/III dominated by the HNL production from the mesons pions/kaons/$D$-$D_s$, respectively, the new sensitivity line would follow the $\um2$ results until $M_N < m_M - m_\mu$, where the phase space for producing the HNL and the muon closes, at which point the $\ue2$ results would dominate the sensitivity.

\subsection{The impact of the background}
\label{subsec:background}

Although a full background analysis would require simulating the propagation of all particles crossing the detector and of their decay products, along with a detector reconstruction; our focus in this paper is on a more qualitative approach. The scope of this section is to estimate the impact of the dominant background on the estimated sensitivity. We believe that the background from cosmic muons and the activity of the rock can be reduced to negligible levels by applying directionality and energy cuts, and thus we consider the light neutrino background to be the dominant one in our analysis. A more sophisticated and comprehensive background analysis should be addressed in a future study.\\
Charged pions cannot be calorimetrically separated from muons in LArTPCs, although they can be identified in cases where they inelastically scatter with an argon nucleus before stopping or exiting the detector. In this work, the conservative approach is taken of treating muons and charged pions as indistinguishable particles ($\mu$-like particle) in the ICARUS detector. Therefore, for the background analysis, it is convenient to categorize the events of the main HNL decay channels discussed in Section~\ref{subsec:sensitivity_channel} in the following way:

\begin{itemize}
    \item Channels with $2\mu$-like particles in the final state. The signal channels are $N\to \mu^+\mu^-\nu$ for the three $\ua2$;  $N\to \pi^+\mu^-$ + $N\to \rho^+\mu^-, \rho^+\to \pi^+\pi^0$ in the particular case of $\um2$, and $N\to \rho^0\nu, \rho^0\to \pi^+\pi^-$ in the particular case of $\ut2$. The main source of background for these channels is single pion production from $\nu_\mu$, and an additional subleading contribution of $\nu_\mu$-$^{40}\text{Ar}$ DIS where the charm meson produces a secondary muon.
    \item Channels with $1\mu$-like particle in the final state. The signal channels are $N\to \mu^\pm e^\mp\nu$ for $\ua2$; and $N\to \pi^+ e^-$ + $N\to \rho^+ e^-, \rho^+\to \pi^+\pi^0$ in the particular case of $\ue2$.  In this category, the main source of background is single pion production from $\nu_e$-$^{40}\text{Ar}$ interactions, and the suppressed contribution of $\nu_\alpha$-$^{40}\text{Ar}$ DIS where the charm meson decays into $\ell_\beta$, with $\alpha,\beta = e,\mu$ and $\alpha\neq\beta$. 
    \item Channels with $0\mu$-like particles in the final state. The signal channels are $N\to e^\pm e^\mp \nu$ and $N\to \pi^0 \nu$ for $\ua2$. With the main background sources coming from single neutral pion production and $\nu_e$-$^{40}\text{Ar}$ DIS where the charm meson decay produces an additional electron.
\end{itemize}

We performed a sensitivity estimate that includes all potential background scenarios for the case of $2\mu$-like final state particles. This highly promising category has been chosen for our analysis because it includes clean decay channels that can be easily identified and reconstructed within the LArTPC detector. 
From the total number of $\nu_\mu$-$^{40}\text{Ar}$ interactions per year listed in Table~\ref{tab:nu_interactions}, $0.12\%$ correspond to events with only $2\mu$-like tracks with energies above threshold in the final state. In this work, a conservative proton detection threshold  of 50~MeV for the ICARUS detector has been considered. Notice that the spectra of the signal and background events are slightly different, due to the fact that the decay products of the HNL are boosted in the direction of the beam. This allows us to maximize the signal over the background by applying a series of kinematical cuts in the energy of the $2\mu$-like tracks ($E_\text{pair}$), in the transverse momentum of the pair (${p_T}_\text{pair}$), and in the angle of $2\mu$-like tracks (${\theta}_\text{pair}$). We found that the expected background could be dramatically reduced by approximately $90\%$ ($\sim 60\%$) for lower to intermediate (higher) $\mn$ values, while keeping in every case more than $\sim 85\%$ of the signal. Table~\ref{tab:signal_background} summarizes the percentage of signal efficiencies and background rejection efficiencies obtained after applying the corresponding event cuts for each of the flavors $\ua2$, and for two different mass regions of $\mn$.

\begin{table}[htb!]
\begin{center}
\begin{small}
\renewcommand{\arraystretch}{1.8}
    \begin{tabular}{| c || c | c |}
    \hline
\multirow{6}{*}{$\ue2$} &  $\mn = 406$~MeV & $\mn = 1020$~MeV \\
    \cline{2-3}
     & $E_\text{pair}<2.9$~GeV & $E_\text{pair}<20.2$~GeV \\
     & ${p_T}_\text{pair}<0.3$~GeV & ${p_T}_\text{pair}<3.2$~GeV \\
     & $\theta_\text{pair}<0.8$~rad & $\theta_\text{pair}<0.3$~rad \\
    \cline{2-3}
     & $\text{Signal eff.}\rightarrow 86.9\%$ & $\text{Signal eff.}\rightarrow 84.3\%$ \\
     & $\text{Bkg. rej. eff.}\rightarrow 89.7\%$ & $\text{Bkg. rej. eff.}\rightarrow 66.9\%$ \\
    \hline
    \hline
\multirow{6}{*}{$\um2$} &  $\mn = 369$~MeV & $\mn = 1020$~MeV \\
    \cline{2-3}
     & $E_\text{pair}<2.9$~GeV & $E_\text{pair}<30.1$~GeV \\
     & ${p_T}_\text{pair}<0.3$~GeV & ${p_T}_\text{pair}<3.2$~GeV \\
     & $\theta_\text{pair}<1.4$~rad & $\theta_\text{pair}<0.4$~rad \\
    \cline{2-3}
     & $\text{Signal eff.}\rightarrow 87.2\%$ & $\text{Signal eff.}\rightarrow 86.7\%$ \\
     & $\text{Bkg. rej. eff.}\rightarrow 86.4\%$ & $\text{Bkg. rej. eff.}\rightarrow 59.6\%$ \\
    \hline
    \hline
\multirow{6}{*}{$\ut2$} &  $\mn = 401$~MeV & $\mn = 1421$~MeV \\
    \cline{2-3}
     & $E_\text{pair}<24.1$~GeV & $E_\text{pair}<22.9$~GeV \\
     & ${p_T}_\text{pair}<2.3$~GeV & ${p_T}_\text{pair}<2.2$~GeV \\
     & $\theta_\text{pair}<0.2$~rad & $\theta_\text{pair}<0.3$~rad \\
    \cline{2-3}
     & $\text{Signal eff.}\rightarrow 91.9\%$ & $\text{Signal eff.}\rightarrow 86.3\%$ \\
     & $\text{Bkg. rej. eff.}\rightarrow 75.5\%$ & $\text{Bkg. rej. eff.}\rightarrow 66.9\%$ \\
    \hline
    \end{tabular}
    \caption{Summary of the kinematical cuts that have been applied to the signal and background events with  $2\mu$-like particles in the final state. For the three flavors of the mixing, two different HNL mass regions have been selected to maximize the sensitivity. In each case, the achieved signal efficiency and background rejection efficiency are shown as a percentage.}
    \label{tab:signal_background}
\end{small}
\end{center}
\end{table}

\begin{center}
    \begin{figure*}[htb!]
        \includegraphics[width=0.86\textwidth]         {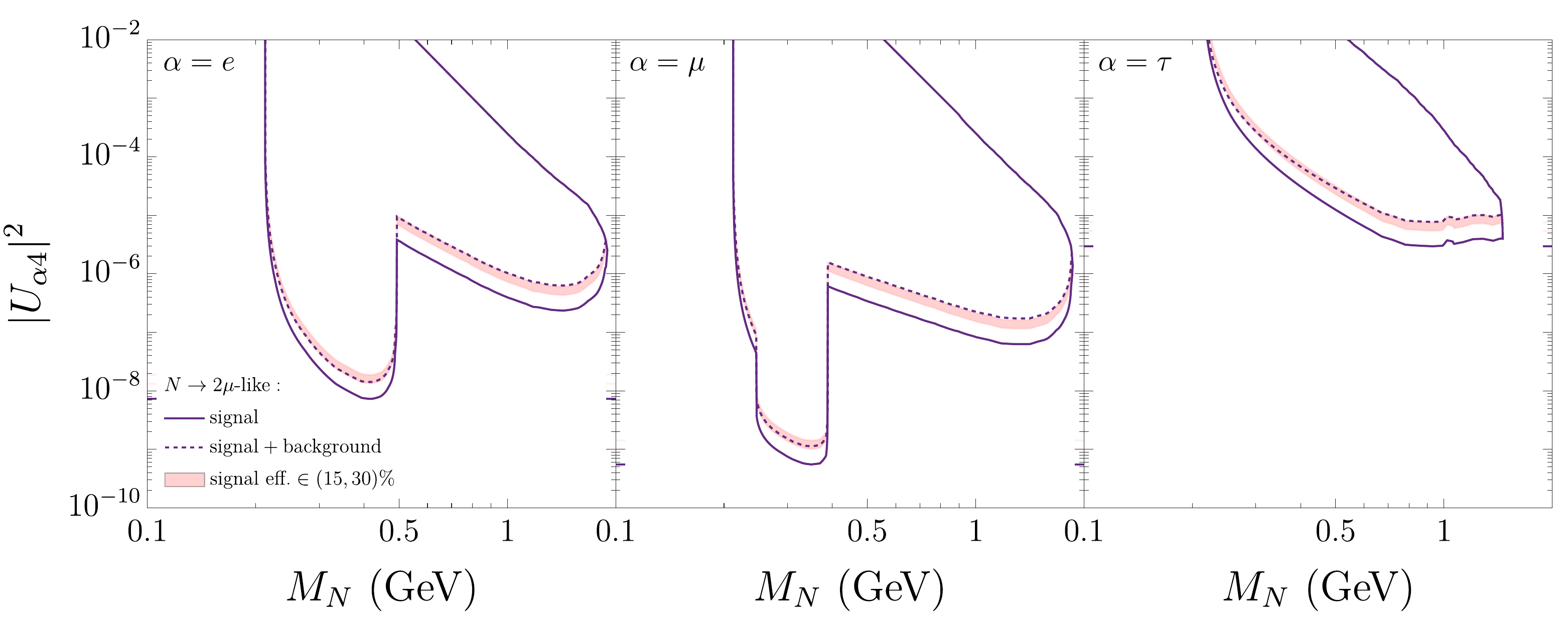}
        \caption{ICARUS-like expected sensitivity for different mixing matrix elements:  $\ue2$ (left panel), $\um2$ (middle panel), and $\ut2$ (right panel) at 90$\%$ C.L., a total collected $1.32 \cdot 10^{21}$ PoT from the NuMI beam, and considering only HNL decays channels with $2\mu$-like final state particles. The sensitivity is presented with signal only (solid-violet), signal +  background (dashed-violet), and signal with an efficiency lying in the $(15, 30)\%$ range (pink-band).} 
        \label{fig:di-muon}
    \end{figure*}
\end{center}

 The ICARUS expected sensitivity for $2\mu$-like final state channel is shown in Figure~\ref{fig:di-muon}. The analysis shows the sensitivity for different mixing matrix elements:  $\ue2$ (left), $\um2$ (middle), and $\ut2$ (right) at 90$\%$ C.L. and an exposure of $1.32 \cdot 10^{21}$ PoT from the NuMI beam. The sensitivity assuming signal only, and obtained by following Eq.~(\ref{eq:Feldman-Cousins}) is presented by a solid violet line. The dashed violet line corresponds to the signal $+$ background analysis following Eq.~(\ref{eq:Feldman-Cousins}) after applying the event cuts for each HNL mass region summarized in Table~\ref{tab:signal_background}. Also, we show that after applying a signal efficiency varying in the range $(15,30)\%$ to the negligible background case, we obtain the pink-band that covers the signal $+$ background result. 

In the case of channels with $1\mu$-like final particles, the main background arises from $\nu_e$-$^{40}\text{Ar}$ interactions; and according to Table~\ref{tab:nu_interactions}, $\nu_e$ interactions from the NuMI beam are one order of magnitude smaller than $\nu_\mu$ ones. This implies that the background will be one order of magnitude suppressed with respect to the $2\mu$-like final particles case. On the other hand, the expected reconstruction efficiency of high-energy electron showers is worse than that of muons tracks in LArTPCs. Thus, we expect the signal to have a sensitivity between $(15, 30)\%$ with negligible background. However, it's important to note that in real experimental data, there will be signal efficiencies due to experimental smearing and other factors, independent of the background.  Finally, in the case of channels with $0\mu$-like final particles, the $\pi^0$ produced either vian HNL decay (signal), or via $\nu_\alpha$-$^{40}\text{Ar}$ NC single $\pi^0$ production, will decay into $2\gamma$ that after propagation in the detector will eventually convert into two $e^+e^-$ pairs. Depending on the energy of the photons, and on the distance the $\gamma$s propagated inside the detector, the $\pi^0$ will produce signatures involving different electron showers. Hence, a detailed simulation of the propagation and $\gamma$ decay followed by a proper detector reconstruction would be needed for a full computation of the expected sensitivity of these channels. Present analyses in the literature (see for instance~\cite{MicroBooNE:2023eef, Coloma:2023oxx}) show that our conservative result of signal efficiency varying between $(15, 30)\%$ could be extrapolated to channels with $0\mu$-like final particles.

\subsection{The potential of ICARUS over present constraints}
\label{subsec:potential_ICARUS}

\begin{center}
    \begin{figure*}[htb!]
        \includegraphics[width=0.86\textwidth]
        {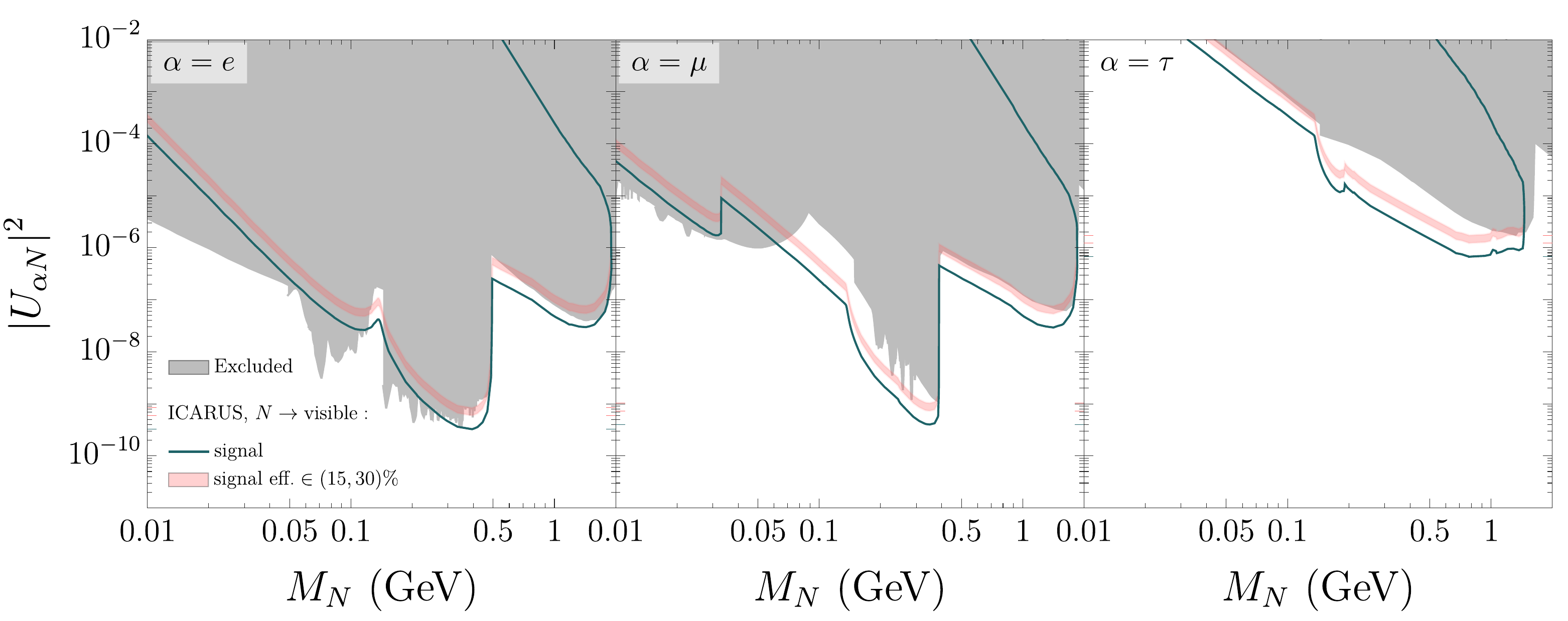}
        \caption{Expected sensitivity at 90$\%$ C.L. of the ICARUS-like detector exposed to $1.32 \cdot 10^{21}$ PoT from the NuMI beam, combining all possible decay channels for the HNL leading to visible final states in the detector. Results are shown for an HNL coupled to $e$ (left panel), $\mu$ (middle panel), and $\tau$ (right panel). Each panel represents the expected sensitivity on the HNL mixing as a function of its mass, with signal only (solid-green), and signal with an efficiency lying between $(15, 30)\%$ (pink-band). The shaded gray areas are disfavored at $90\%$ C.L. by present experiments searching for Dirac HNLs have been obtained with \texttt{HNLimits}~\cite{Fernandez-Martinez:2023phj}. The complete list of current experiments setting the most stringent constraints on HNL mixing is provided at the end of Section~\ref{sec:HNL_pheno}.}
        \label{Fig:summary_plot}
    \end{figure*}
\end{center}

Figure~\ref{Fig:summary_plot} summarizes the ICARUS expected sensitivities at 90$\%$ C.L. to the heavy neutrino mixing $\ua2$ as a function of its mass, for  $1.32 \cdot 10^{21}$ PoT from the NuMI beam. The results combine all the possible HNL decay channels leading to visible final states in the detector, as discussed in Section~\ref{subsec:sensitivity_channel}, and shown in Figure~\ref{fig:signal_only}. The combined sensitivity Figure~\ref{Fig:summary_plot} is presented with signal only (solid dark green), and signal with an efficiency lying between $(15, 30)\%$ (pink band) following the Feldman and Cousins prescription of Eq.~(\ref{eq:Feldman-Cousins}). The shaded gray areas represent the regions of the parameter space that have been already excluded by present experiments, discussed in Section~\ref{sec:HNL_pheno}. The \textsc{GitHub} package and repository \texttt{HNLimits}~\cite{Fernandez-Martinez:2023phj} has been used to adjust the present constraints to the same 90\% C.L., and the same nature assumed for the neutrinos. 

The ICARUS detector has the potential to extend the search for HNLs in  parameter space not yet explored by present experiments. In Table~\ref{tab:ICARUS_limits}, the possible HNL mass windows and the expected lowest reachable 
mixing value of the sensitivity of the corresponding mixing matrix element are summarized as potential upper bounds set by the experiment. In the case of $\ut2$, the lower value of the HNL mass window corresponds to the intersection of our result and the exclusion region from Borexino~\cite{Plestid:2020ssy}. This crossing is not shown in Figure~\ref{Fig:summary_plot} since we chose to plot the range $\ut2>10^{-2}$.

\begin{table}[h]
\begin{center}
\begin{small}
\renewcommand{\arraystretch}{1.8}
    \begin{tabular}{| c || c | c | c |}
    \hline
 & & \multicolumn{2}{c|}{$\ua2 \gtrsim$} \\
\cline{3-4}
 & $\mn/\text{MeV}\in$ & ICARUS & Present bound \\
    \hline
    \hline
   \multirow{2}{*}{$\alpha = e$}  & (122, 144) & $2.3\cdot 10^{-8}$ & \JE{$1.4\cdot 10^{-7}$} \\
     & $(\JE{350}, 1630)$ & $3.3\cdot 10^{-10}$ & \JE{$5.5\cdot 10^{-10}$} \\
    \hline
  \multirow{1}{*}{$\alpha = \mu$} & (\JE{62}, 1862) & $4.0\cdot 10^{-10}$ & \JE{$1.2\cdot 10^{-9}$} \\
    \hline
\multirow{1}{*}{$\alpha = \tau$} & $(\JE{14}, 1450)$ & $6.7\cdot 10^{-7}$ & \JE{$1.8\cdot 10^{-6}$} \\
    \hline
    \end{tabular}
    \caption{HNL mass regions where ICARUS-like detector is expected to test parts of the parameter space unexplored by present experiments. For each mass region, the middle (right) column shows the lowest point of the ICARUS (present) sensitivity is indicated as the upper bound of the corresponding mixing matrix element $\ua2$.}
    \label{tab:ICARUS_limits}
\end{small}
\end{center}
\end{table}

\section{Summary}
\label{sec:conclusions}

One of the simplest ways to accommodate the observed pattern of neutrino masses and mixings is the addition of heavy right-handed neutrinos, or HNLs, to the SM particle content. The mass of the HNLs is a dimensionful free parameter of the model, and experimental verification is needed for its determination. When the mass of the HNL ranges the MeV-GeV scale, and for favorable 
mixing scenarios, present and near-future beam-dump experiments set a perfect environment for their discovery.

In this paper, we present a search for HNL using the NuMI beam data corresponding to $1.32\cdot 10^{21}$ PoT with the ICARUS detector.
In our analysis, we interpret the results using three different sets of figures. We first compare the sensitivity of different HNL decay channels across a wide mass range (see Figure~\ref{fig:signal_only}).
Next, we demonstrate the impact of the dominant background on one of the clean and more straightforward signal channels at ICARUS, when the HNL decays into a pair of muon-like particles (see Figure~\ref{fig:di-muon}). The efficiency obtained from the background simulation of this signal channel is comparable to uncovering the signal efficiency range for other promising channels, such as $N\to e^-e^+\nu$ and $N\to \pi^0\nu$, as extensively studied in the literature~\cite{MicroBooNE:2023eef, Coloma:2023oxx}. Hence, we have used the signal efficiency region between $(15, 30)\%$ for generating sensitivity for all other promising signal channels.\\

This information will be valuable for future analyses and will also provide us with an estimate of the signal efficiency for the analysis assuming negligible background. Finally, we present the sensitivity results with signal only, and with varying signal efficiency, to emulate a negligible-background analysis (see Figure~\ref{Fig:summary_plot}), where we also compare the results with the leading constraints from present experimental data listed at the end of Section~\ref{sec:HNL_pheno}. We show that ICARUS could improve over present data in certain HNL mass regions, depending on the flavor of the mixing that is assumed (see Table\ref{tab:ICARUS_limits}). In case 
the HNL mixes only with electrons, there are two unexplored mass windows that ICARUS could test: $\mn\in (122, 144)$~MeV with a peak sensitivity of $\ue2\gtrsim 2.3\cdot 10^{-8}$, and $\mn\in (350, 1630)$~MeV with an upper bound of $\ue2\gtrsim 3.3\cdot 10^{-10}$. Furthermore, ICARUS could test an HNL 
that only mixes with muons  if $\mn\in (62, 1862)$~MeV, setting a upper bound on its mixing $\um2\gtrsim 4.0\cdot 10^{-10}$. In the case of the mixing with the $\tau$, where the present bounds are lest stringent, ICARUS could improve the bound in the full accessible HNL mass range $\mn\in (\JE{14}, 1450)$~MeV to $\ut2\gtrsim 6.7\cdot 10^{-7}$. Note that in the tau mixing case, the intersection between our sensitivity line and the current bound from Borexino~\cite{Plestid:2020ssy}, which happens at $M_N \simeq 14$~MeV, is not shown in Figure~\ref{Fig:summary_plot} because it occurs at $\ut2\simeq 0.13$, while we chose to plot the range $\ut2<10^{-2}$. In addition, the ICARUS experiment has the ability to discover HNL by measuring the invariant mass of the decay channels. The discovery of HNL can be done by measuring invariant mass of the two muon like final state (exclusive channel like fully contained $\mu-\pi$ or $\mu-\mu$). This is one of the main motivation of our paper to show that 2 muon like channel  can be important to look by the ICARUS collaboration.\\
The SBN proposal states that ICARUS will receive $6.6 \cdot 10^{20}$ PoT per year,
thus our result  corresponds to 2 years of ICARUS data collection. However, a recent presentation by the ICARUS collaboration at the Neutrino 2024 conference showed that the ICARUS experiment has gathered $5.44 \cdot 10^{20}$ PoT during
the first two years (including neutrino and anti-neutrino mode). Therefore, the experiment is expected to achieve the desired sensitivity in the next two and a half years with
a similar data rate collection as in 
previous years.

The results of this HNL sensitivity study 
may open the door for future searches at ICARUS, covering all potential HNL decay channels, as well as a wider range of BSM measurements that are part of the SBN program.

\acknowledgments
    We warmly thank Sharoz Schezwen for collaborating during early stages of this work, and Zarko Pavlovic for fruitful discussion regarding the simulations of the heavy meson. We also acknowledge the SOM cluster at the IFIC, where all the simulations regarding this work have been performed. AC acknowledges CERN Neutrino Platform and the Ramanujan Fellowship (RJF/2021/000157), of the Science and Engineering Research Board of the Department of Science and Technology, Government of India. The research of JHG is supported by the EU H2020 research and innovation programme under the MSC grant agreement No 860881-HIDDeN, as well as by the Spanish Ministerio de Ciencia e Innovaci\'on project PID2020-113644GB-I00, the Spanish Research Agency (Agencia Estatal de Investigaci\'on) through the project CNS2022-136013, and Severo Ochoa Excellence Program CEX2023-001292-S.
    JHG thanks the CERN EP Neutrino group for hospitality. 
    ADR acknowledges CERN Neutrino platform.
    It is to be noted that this work has been done solely by the authors and is not representative of the ICARUS collaboration.

\appendix

\section{Light neutrino flux prediction and its cross section with $^{40}$Ar.}
\label{app:flux}

\begin{center}
    \begin{figure*}[htb!]
        \includegraphics[width=0.4\textwidth]
        {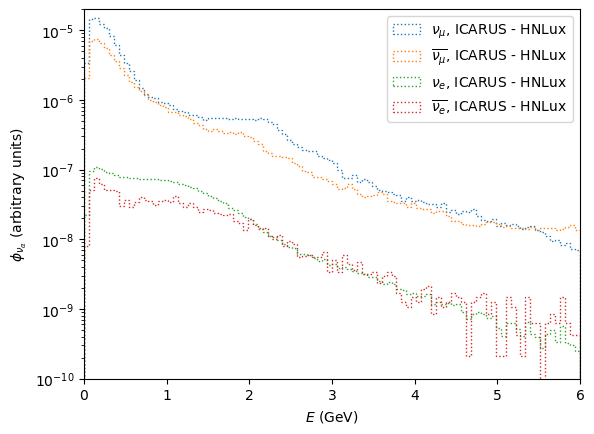}
        \includegraphics[width=0.4\textwidth]
        {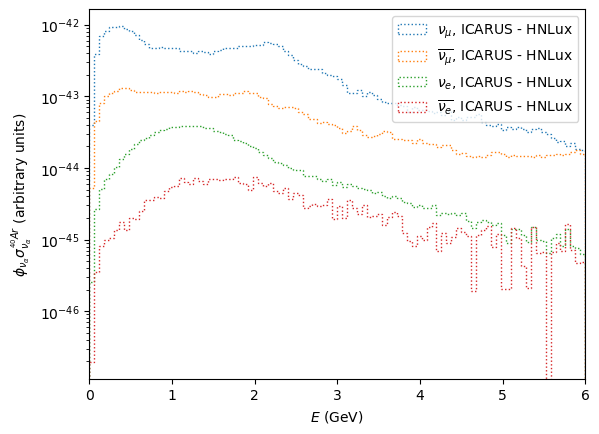}
        \caption{Left panel: Prediction of the light neutrino fluxes crossing the geometry of the ICARUS detector. The different colors mean different flavor of the electron and muon neutrinos and antineutrinos. Right panel: convolution of the fluxes and the $\nu_\alpha$-$^{40}$Ar cross sections taken from GENIE.}
        \label{fig:flux}
    \end{figure*}
\end{center}

Starting from the G4NuMI simulations containing the parent particles produced at the NuMI target, we used \texttt{HNLux} to compute the expected fluxes of the light neutrinos and antineutrinos crossing the geometry of the ICARUS detector. The left panel of Figure~\ref{fig:flux} shows the energy distribution of these fluxes. The main contributions to the electron neutrino flux come from $\mu^\pm$ decay as well as the semileptonic 3-body decays of $K^\pm$ and $K^0_L$. In the case of muon neutrinos, the flux is mostly dominated by the 2-body decays of $\pi^\pm$ and $K^\pm$. Our results have been validated against the flux prediction as shown by the ICARUS collaboration.
The number of neutrino interactions with the $^{40}$Ar in the ICARUS detector is determined by the convolution of the flux and the cross section, weighted by the number of targets inside the detector volume. The right panel of Figure~\ref{fig:flux} shows the convolution of the four fluxes and the NN and CC cross sections of $\nu_\alpha$-$^{40}$Ar as provided by GENIE. Specifically, the precompiled cross section G1802a00000-k250-e1000 from GENIE v3.04.00 was used for this plot.

\section{Kinematical cuts}
\label{app:cuts}

\begin{center}
    \begin{figure*}[htb!]
        \includegraphics[width=0.32\textwidth]
        {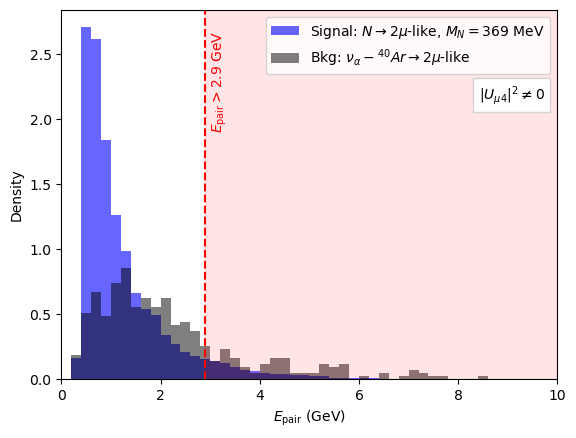}
        \includegraphics[width=0.32\textwidth]
        {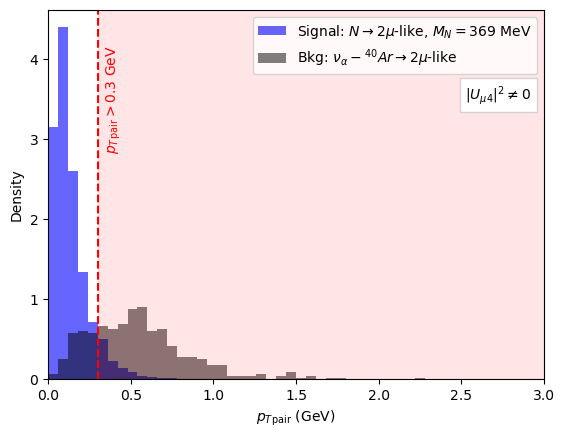}
        \includegraphics[width=0.32\textwidth]
        {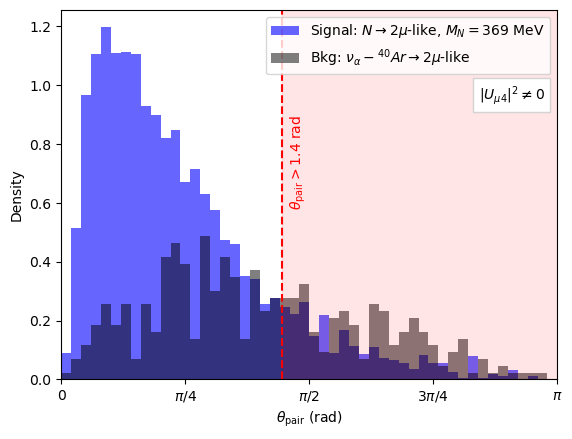}    
        \includegraphics[width=0.32\textwidth]
        {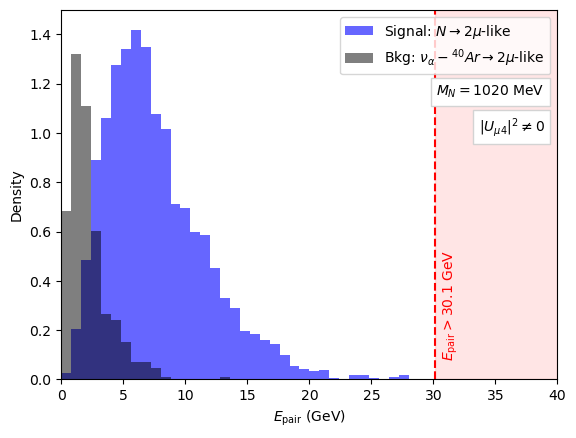}
        \includegraphics[width=0.32\textwidth]
        {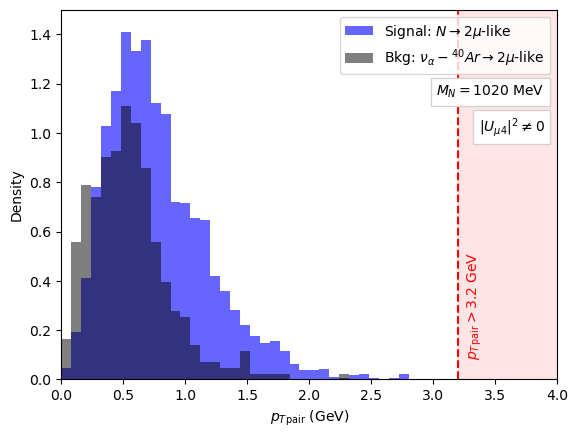}
        \includegraphics[width=0.32\textwidth]
        {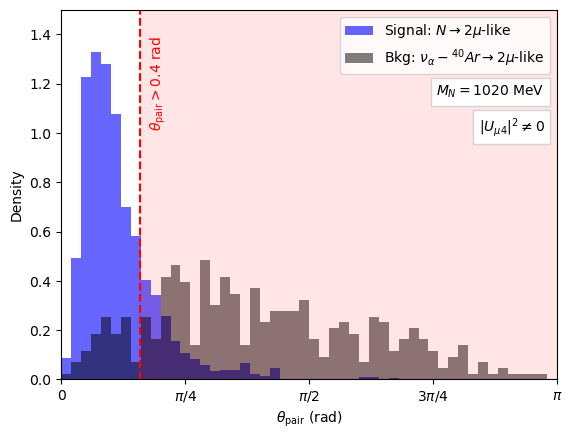}
        \caption{Signal (blue histograms) and background (gray histograms) distributions of events in the kinematical variables $E_\text{pair}$ (left panels), ${p_T}_\text{pair}$ (middle panels), and $\theta_\text{pair}$ (right panels). The upper panels correspond to an HNL of 369~MeV, while the lower panels correspond to an HNL of 1020~MeV. The shaded areas in red represent the events that are removed after each kinematical cut.}
        \label{fig:kinematical_cuts}
    \end{figure*}
\end{center}

The signal and background follow dissimilar distributions of the kinematical variables $E_\text{pair}$, ${p_T}_\text{pair}$, and $\theta_\text{pair}$. Thus, applying different kinematical cuts allows us to reduce the background while maintaining maximal signal, as discussed in Section~\ref{subsec:background}. Figure~\ref{fig:kinematical_cuts} shows the distributions of signal (blue histograms) and background (gray histograms) with the event selection of $2\mu$-like particles in the final state. The upper panels correspond to the case of $M_N = 369$~MeV, while the lower panels correspond to $M_N = 1020$~MeV; in both cases mixed with muons only. This means that in the upper panels, the HNL has been produced from kaon decays, whereas in the lower panels, it comes from $D/D_s$ decays. As explained in Section~\ref{sec:signal_background}, the production of these mesons in NuMI is very different. While the light mesons are mostly produced at the target station and focused by the horn system placed immediately afterward, the heavy mesons are produced both at the target and at the absorber at the end of the decay pipe and decay promptly without being focused. This results in dissimilar distributions of the signal events in the two mass hypotheses; see, for instance, the ${p_T}_\text{pair}$ distributions in the middle panels of Figure~\ref{fig:kinematical_cuts}. Consequently, the achieved signal efficiency and background rejection efficiency after the kinematical cuts for the two situations are very different, as seen in Table~\ref{tab:signal_background}.

\bibliographystyle{apsrev4-1}
\bibliography{references}

\end{document}

%% file: custom_comands.tex
\def\ua2{\vert U_{\alpha N}\vert^2}
\def\ue2{\vert U_{e N}\vert^2}
\def\um2{\vert U_{\mu N}\vert^2}
\def\ut2{\vert U_{\tau N}\vert^2}
\def\mn{M_N}